\documentclass[aps,reprint,pra, twocolumn,superscriptaddress,floats,floatfix]{revtex4-1}

\usepackage{siunitx}
\usepackage{amsmath}
\usepackage{amsfonts}
\usepackage{amssymb}
\usepackage{graphicx}
\usepackage{color}
\usepackage[colorlinks=true,citecolor=blue,linkcolor=blue,urlcolor=blue]{hyperref}
\usepackage{times}
\usepackage{amstext}
\usepackage{latexsym}
\usepackage[running,mathlines]{lineno}
\usepackage{verbatim}
\usepackage{bm}
\usepackage[version=3]{mhchem}

\usepackage{cleveref}

\providecommand{\ket}[1]{\lvert #1 \rangle}

\providecommand{\ketbra}[2]{\lvert  #1\rangle \langle #2 \rvert}

\begin{document}
\raggedbottom

\title{ Optimizing resetting of superconducting qubits}

\author{Ciro Micheletti Diniz}
\address{Departamento de Física, Universidade Federal de São Carlos, 13565-905,
São Carlos, São Paulo, Brazil}
\author{Rogério Jorge de Assis}
\address{Departamento de Física, Universidade Federal de São Carlos, 13565-905,
São Carlos, São Paulo, Brazil}
\author{Norton G. de Almeida}
\address{Instituto de Física, Universidade Federal de Goiás, 74.001-970, Goiânia
- GO, Brazil}
\author{Celso J. Villas-Boas}
\address{Departamento de Física, Universidade Federal de São Carlos, 13565-905,
São Carlos, São Paulo, Brazil}

\begin{abstract}
    Many quantum algorithms demand a large number of repetitions to obtain reliable statistical results. Thus, at each repetition it is necessary to reset the qubits efficiently and precisely in the shortest possible time, so that quantum computers actually have advantages over classical ones. In this work, we perform a detailed analysis on three different models for information resetting in superconducting qubits. Our experimental setup consists of a main qubit coupled to different auxiliary dissipative systems, that are employed in order to perform the erasing of the information of the main qubit. Our analysis shows that it is not enough to increase the coupling and the dissipation rate associated with the auxiliary systems to decrease the resetting time of the main qubit, a fact that motivates us to find the optimal set of parameters for each studied approach, allowing a significant decrease in the reset time of the three models analyzed.
\end{abstract} 
\maketitle

\section{Introduction}
Beyond the preparation with a high fidelity of the qubit initial states \cite{Kleiler2018,Huang2019} and the possibility of implementing error correction algorithms \cite{Fowler2012,Dennis2002,Gottesman1997,Wang2003}, a key step in building an operational quantum computer is the capacity of making these devices process data as quickly as possible. Due to the probabilistic character of the Quantum Mechanics, many quantum algorithms demand a large number of repetitions and, consequently, the resetting of the qubits must be repeated many times. In this sense, a strategy to speed up the computing is the optimization of the reset process through the elaboration of a fast way to make the qubits return to their initial states. To illustrate the significance of reinitializing the system in the entire process of running an algorithm, in Ref. \cite{Xin_2020} a set of $N$ coupled linear differential equations is solved and the solution is mapped to the final state of $n$ qubits (with $N=2^n$), such that, to achieve a precise solution, it is necessary to repeat the measurements up to $\mathcal{O}(2^n)$ times, since there are $2^n$ possible states for the qubits. For instance, we estimate that to solve a set with one trillion of coupled differential equations it would be need to repeat the measurements around one trillion times. In this process, if we consider a realistic reset time as in Ref. \cite{IBM,Barrett2013}, which is of the order of hundreds of nanoseconds or even a few microseconds, it would take approximately some tens or even a few hundreds hours in the re-initialization steps of the system. In order to create a faster resetting, several approaches can be applied. One of them consists in performing measurements on the qubits to invert their states using controlled duration pulses \cite{Rist2012,Blais2004,Wallraff2004,Reed2010}. Other ways to produce an efficient reset process use thermal baths to stabilize the system in a specific final state \cite{Tuorila2017, Tuorila2019} or engineer light matter interactions either by pulses with a specific frequency interval \cite{Chen2021} or employing additional level and coupling the qubit to auxiliary dissipative cavity mode \cite{IBM}.

Looking at the fastest way to reset systems, we analyze here two configurations that use dissipative effects to reset the qubit. In addition, we optimize the reset model from Ref. \cite{IBM} and compare its results with the ones achieved with the two new models. The different experimental setups, which will be explored throughout this manuscript, are indicated in Fig. \ref{New Schemes}(a)-(e). The first configuration studied here, Fig. \ref{New Schemes}(c), is a work qubit (named as main qubit) coupled directly to a second highly dissipative qubit (auxiliary qubit). In Refs. \cite{Basilewitsch2017, Fischer2019}, it was used a similar model to study how to speed up the resetting and the purification of the system, but differently from the present work, the authors employed a tunable frequency qubit as the main system and considered the presence of correlations between the qubit and the environment. Also, in \cite{Basilewitsch2021}, using a model similar to the previous one, it was studied how the increasing of the size of the ancilla qubit Hilbert space affects the resetting time. In the second configuration, Fig. \ref{New Schemes}(d), the main qubit is coupled to a second one, which is then coupled to a dissipative cavity mode. In \cite{Basilewitsch2019} it was also used two auxiliary components as in our model, but, once again, it was considered tunable frequencies in order to maximize the efficiency of the resetting. Finally, in the third experimental setup investigated here we take into account the multilevel structure of a superconducting device, \textit{i.e.}, in addition to the two levels that work out as the main qubit, an auxiliary third level is considered, which is used to couple the quantum device to a dissipative bosonic field mode, Fig. \ref{New Schemes}(e). This resetting mechanism is quite interesting for its simplicity and efficiency and it was studied before in Refs. \cite{Magnard2018, IBM}, where, in the both of them, the authors employed a Jaynes-Cummings-type interaction to reset the system, but with different approaches, which made the resetting time achieved in the latter to be more than twice as fast as in the former. Still concerning the last configuration, besides of saving space in the quantum computer by eliminating the need for a auxiliary qubit, this scheme can be applied to fixed frequency qubit architectures, as the case of the IBM systems in Refs. \cite{IBM,IBM_Q_EXP}.

%\textcolor{magenta}{To pave the way for the search of an optimal reset protocol applicable to existing quantum computers,} 
In this work we carry out a numerical analysis of the main qubit dynamics for the three configurations cited above and then we perform an optimization for each setup, searching for the set of parameters that makes the resetting process faster. To present our results, this manuscript is organized as follows: Section II introduces the three models shown in Fig. \ref{New Schemes} and it describes two different approaches that will be used in the information erasing, which are: (i) the pulsed and (ii) the steady-state approaches. In Section III, it is shown the parameters that optimize the resetting for each model using these two approaches and it is discussed their experimental feasibility. Finally, Section IV presents our conclusions.

\section{Description of the models}\label{Description of the models}

\begin{figure*}
\begin{center}
\includegraphics{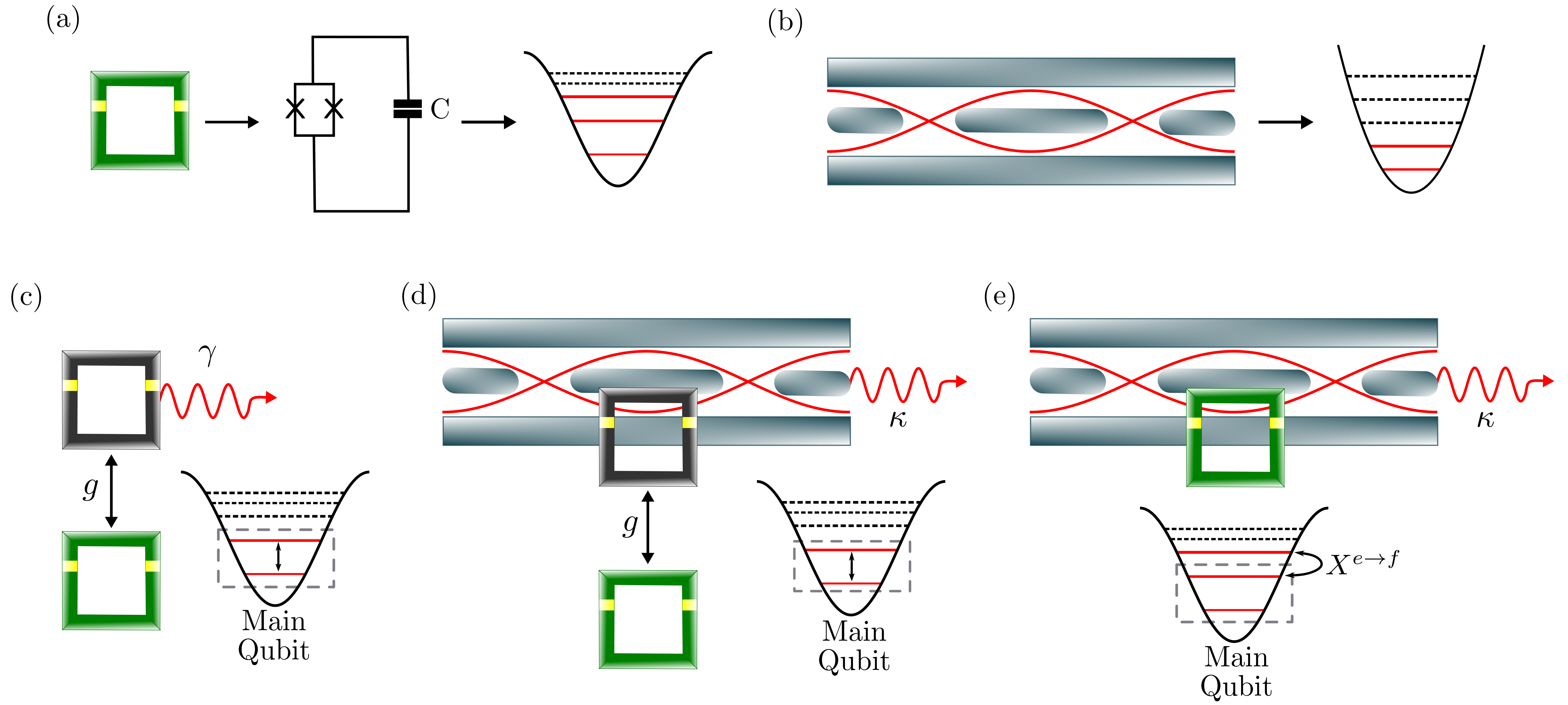}

\caption{Pictorial representation of the proposed schemes for resetting the work qubit. Panel (a) shows a superconducting device, our main qubit, its equivalent circuit composed by a capacitor (C) and a Josephson junction (JJ), which works as a nonlinear inductor, and the respective anharmonic potential with the corresponding energy levels associated with the circuit. Panel (b) shows the waveguide (cavity mode) with the corresponding harmonic potential and energy levels. Panel (c) shows the coupling between the main (green) and the dissipative (black) auxiliary qubit, with coupling strength $g$. The dissipative qubit has damping rate $\gamma$. Panel (d) shows the main qubit coupled to the auxiliary one, with coupling strength $g$. The auxiliary qubit is then coupled (coupling strength $\lambda$) to a dissipative cavity mode, whose dissipation rate is $\kappa$. Panel (e) shows the qubit with an auxiliary level, which is coupled to the dissipative bosonic mode with damping rate $\kappa$.}\label{New Schemes}
\end{center}
\end{figure*}

To model the resetting mechanism we resort to the open quantum system treatment, working in a regime where the interaction energy between the subsystems that make up our setup is much smaller than the free energy of the subsystems themselves. On the other hand, the coupling strengths between the subsystems can be either stronger or weaker than the their dissipation rates, usually named as strong or weak coupling regimes, respectively. In these regimes, and assuming Born and Markov approximations, the dynamics of the system is governed by the master equation ($\hbar =1$) \cite{breuer2002theory}
\begin{equation}\label{general master equation}
    \frac{\partial \rho}{\partial t} = -i\left[H, \rho \right] + L_F\left(\rho \right) + L_A\left(\rho \right),
\end{equation}
in which the Hamiltonian $H$ describes the system coupled to the auxiliary components,  $\rho$ is the density operator of the composed setup, $L_F\left(\rho \right)=\frac{\kappa}{2}\left(2 a\rho a^{\dagger} - a^{\dagger}a\rho -\rho a^{\dagger}a \right)$ describes the dissipation in a quantum bosonic mode and $L_A\left(\rho \right)=\frac{\gamma}{2}\left(2 \sigma^A_-\rho \sigma^A_+ - \sigma^A_+\sigma^A_-\rho - \rho \sigma^A_+\sigma^A_-\right)$ describes the energy loss of an ancilla qubit. The decay rate of the auxiliary qubit and the bosonic mode are given by $\gamma$ and $\kappa$, respectively, $\sigma^A_+$ ($\sigma^A_-$) is the Pauli raising (lowering) operator for the auxiliary qubit  and $a$ ($a^{\dagger}$) is the annihilation (creation) operator for the dissipative field mode. Since Eq. (\ref{general master equation}) is a general equation, it can describe the three configurations studied here, therefore the Hamiltonian $H$ and the presence of the dissipation operators in this equation will depend on the model, which will be detailed at the moment when we formally introduce each setup. Once we know the configuration that will be analyzed, we are able to replace in Eq. (\ref{general master equation}) the corresponding Hamiltonian that will rule the dynamics in each case and the respective dissipation operators of the auxiliary components and, then, proceeding with the simulations. At this point, one can notice the lack of dissipation operators for the main qubit, our system. In this paper, its absence is a reasonable assumption because quantum computing requires qubits with negligible decay rates, thus we are considering the main qubit as a non-dissipative system, hence the entire process of energy dissipation takes place via auxiliary components.

The three configurations for resetting the main qubit that we are going to study here are pictorially represented in Fig.\ref{New Schemes}. In panel (a) we represent a superconducting device whose the two lower levels represent our main qubit. In panel (b) we represent the waveguide, which works as a dissipative cavity mode and it is described by a harmonic potential. In panel (c) we represent the two-qubit model, where the main qubit is coupled to an auxiliary dissipative one. In panel (d), we represent the main qubit coupled to a second one, which in turn is coupled to a dissipative bosonic mode. Finally, in panel (e), we display the system studied in Ref. \cite{IBM}, which is composed by the main qubit and the auxiliary third level that is coupled to a dissipative cavity mode.

To start, let us introduce the first configuration, as shown in Fig. \ref{New Schemes}(c), called here by two-qubit model. This setup contains the main qubit coupled to a dissipative auxiliary one and has the advantage of being very simple and compact, saving space in a quantum computer. Also, the auxiliary qubit does not need to be fully controllable, since we do not perform operations over its states. On the other hand, as it can emit in any direction, its dissipation could affect the work qubits in a huge quantum computer, depending on the configuration. Besides, to avoid extra noise and undesired dissipation in the main qubit during the execution of a given quantum process, we must be able to turn on and off its interaction with the auxiliary qubit. This can be done assuming that the auxiliary qubit is frequency tunable, being most of the time very far from resonance with the main qubit. Then, to implement the reset process we turn its frequency on resonance with the main qubit, which can be done very quickly, below one nanosecond \cite{Changkang_email}, so we are disregarding this time in our analyses. Thus, during the reset process we can consider these qubits as resonant ones, whose Hamiltonian in the interaction picture  is given by
\begin{equation}\label{hamiltoniam two qubits}
    H_{Q_M-Q_A}= g(\sigma^M_{eg}\sigma^A_{ge} + h.c.),
\end{equation}  
where $g$ is the coupling strength between the main ($M$) and the auxiliary ($A$) qubits, and $\sigma^K_{ij} \equiv \ketbra{i}{j}$ is the transition level operator from level $j$ to level $i$ ($i,j=g,e$) for qubit $K$ ($K= M$,$A$).

The second configuration analyzed here, named two qubits-cavity model, is shown in Fig. \ref{New Schemes}(d) and it is composed by the main qubit coupled to a second non-dissipative and tunable (as in the previous case) auxiliary qubit, which in turn is coupled to a dissipative cavity mode field whose dissipation rate is $\kappa$. In this model, when turning the auxiliary qubit on resonance with the main one and the cavity mode, we end up with the coupling strength $g$ between the two qubits, which can be different from the coupling strength $\lambda$ between the auxiliary qubit and the cavity mode field. In this case, in the interaction picture, the Hamiltonian reads
\begin{equation}\label{H our}
   H_{Q_M-Q_A-F} = g\sigma^M_{eg}\sigma^A_{ge} + \lambda\sigma^A_{eg}a + h.c.,
\end{equation}
where  $a$ ($a^\dagger)$ is the annihilation (creation) operator for the mode and, again, $M$ and $A$ stand for main and auxiliary qubits, respectively. This model requires more elements and space, but the dissipative mode can direct the dissipated energy, thereby avoiding disturbances to neighboring working qubits in a quantum computer.

The Fig. \ref{New Schemes}(e) shows the third model studied here, where the main qubit has an auxiliary level that is coupled to a dissipative bosonic mode. This configuration is named here as IBM model since it was studied by IBM researchers Ref. \cite{Pechal2014,IBM, Zeytinolu2015}. The Hamiltonian that describes this setup, in the rotating frame of the driving field, which has a frequency $\omega_d$, is
\begin{multline}
       H_{IBM} = \delta_c a^\dagger a  + \delta_q b^\dagger b + \frac{\alpha}{2}b^\dagger b^\dagger b b \\
       +( \Tilde{g} a b^\dagger + \frac{1}{2}\Omega(t)b^\dagger  + h.c.),
\label{H ibm}
\end{multline}
where $a$ ($a^\dagger$) is the cavity-mode field annihilation  (creation) operator as in the previous case, $b$ ($b^\dagger$) is the annihilation  (creation) operator of the main system, now considered as a multilevel structure, $\Tilde{g}$ is the coupling strength between the cavity and the system, $\Omega(t)$ is the time-dependent amplitude of the microwave field that drives the qubit, $\delta_c = \omega_{c}-\omega_{d}$,  $\delta_q = \omega_{ge}-\omega_{d}$, with $\omega_{c}$ ($\omega_{ge}$) being the transition frequency of the cavity mode (qubit) and, at last, $\alpha$ is the anharmonicity that modifies the frequency transition $\omega_{ef}$ between the excited state $\ket{e}$ of the qubit and the auxiliary level $\ket{f}$, such that $\omega_{ef} = \omega_{ge} -\alpha$. 

In contrast to the first two configurations cited before, in which the reset processes start when the main qubit begins its interaction with the other components, the resetting in IBM model occurs in two steps: (1) the state population initially in $\ket{e}$ is transferred by an electromagnetic pulse to the ancilla level $\ket{f}$; (2) then the population of the state $\ket{f}$ is transferred to a dissipative cavity mode via its coupling to the transition $\ket{g} \leftrightarrow \ket{f}$ with the help of a driving field with amplitude $\Omega(t)$ and frequency $\omega_d$. In the first step, the time to transfer the population depends on the intensity of the pulse, \textit{i.e.}, its Rabi frequency, such that as higher the Rabi frequencies is, as smaller the time to reach the desired final state. However we cannot increase the intensity indiscriminately, since levels higher than $\ket{f}$ could be populated, in a way that would make the step (2), and consequently the entire reset process, ineffective. To avoid this loss of efficiency, we will consider here the same pulse that was used in Ref. \cite{IBM} to transfer the population to the auxiliary level, which makes the first step of the approach finishes in $75$ $ns$. Hence, in order to speed up the resetting for this configuration, we will focus on optimizing the second step by searching the parameters that dissipate the energy faster.

Besides the three configurations, we also deal with two approaches. In the first, here called the steady-state approach, the reset occurs when the system reaches the ground state in a stable way. In the second approach, which we will call by the pulsed approach following Ref. \cite{IBM}, the reset occurs when the interaction dynamics takes the main qubit to the ground state for the first time and, exactly at that moment, the pulse is turned off and the information is erased. The IBM model was developed based only on a pulsed approach. As explained before, the model studied in Ref. \cite{IBM} involved two steps. On the other hand, for the other two models studied here, the pulsed approach happens just in one step, since we do not have to transfer the population to another level as in the IBM model. Hence, for the other two models, in the pulsed approach the reset process starts when the main qubit is coupled to the auxiliary dissipative systems and it is finished when the required ground state population is achieved for the first time. The time to couple (decouple) the main qubit to (from) the auxiliary components is negligible in the analyses of the reset time, as discussed above.

Now, given the models in Fig. \ref{New Schemes}, we can proceed with the calculations. For comparison purposes, in all simulations we considered the main qubit initially in the excited state, while  all the other auxiliary components were considered to be in their respective ground states. In fact, considering the auxiliary components to be in their ground states is a realistic assumption according to \cite{Chu2022}, since, following their parameters for the mode transition frequency and temperature, the fidelity between the thermal state and ideal ground state is above $99\%$, which is a small error in the state preparation. Next, we solve Eq. (\ref{general master equation}) numerically in Python using the QuTip toolbox \cite{JOHANSSON2012,JOHANSSON2013} considering the  particularities of each model and analyze the dynamics of the ground state population of the main qubit. The study of this property of the main system allows us to determine how and when the reset process must happen.

\section{Results and discussions}

As we are looking at the resetting process, the baseline studied here will be the ground state population of the main qubit. In this way, for the steady-state approach, we consider the reset process performed when the population of the main qubit reaches the minimum value $p_g = 0.98$ without recurring anymore. Usually, to reset properly, quantum computers require a much higher population in the ground state. However, we decided to choose this value in order to perform a fair comparison of our results with the ones presented in \cite{IBM}, which use this population value for the ground state. In the same way, for the pulsed reset approach, we will consider that the process ends when the population of the main qubit reaches the minimum value of $p_g=0.98$ for the first time, which is the instant when the pulse must be turned off. Still, the required time to reach the desired ground state population, the reset time, is named here $t_{stop}$.

\begin{figure}
\begin{tabular}{cc}
\includegraphics{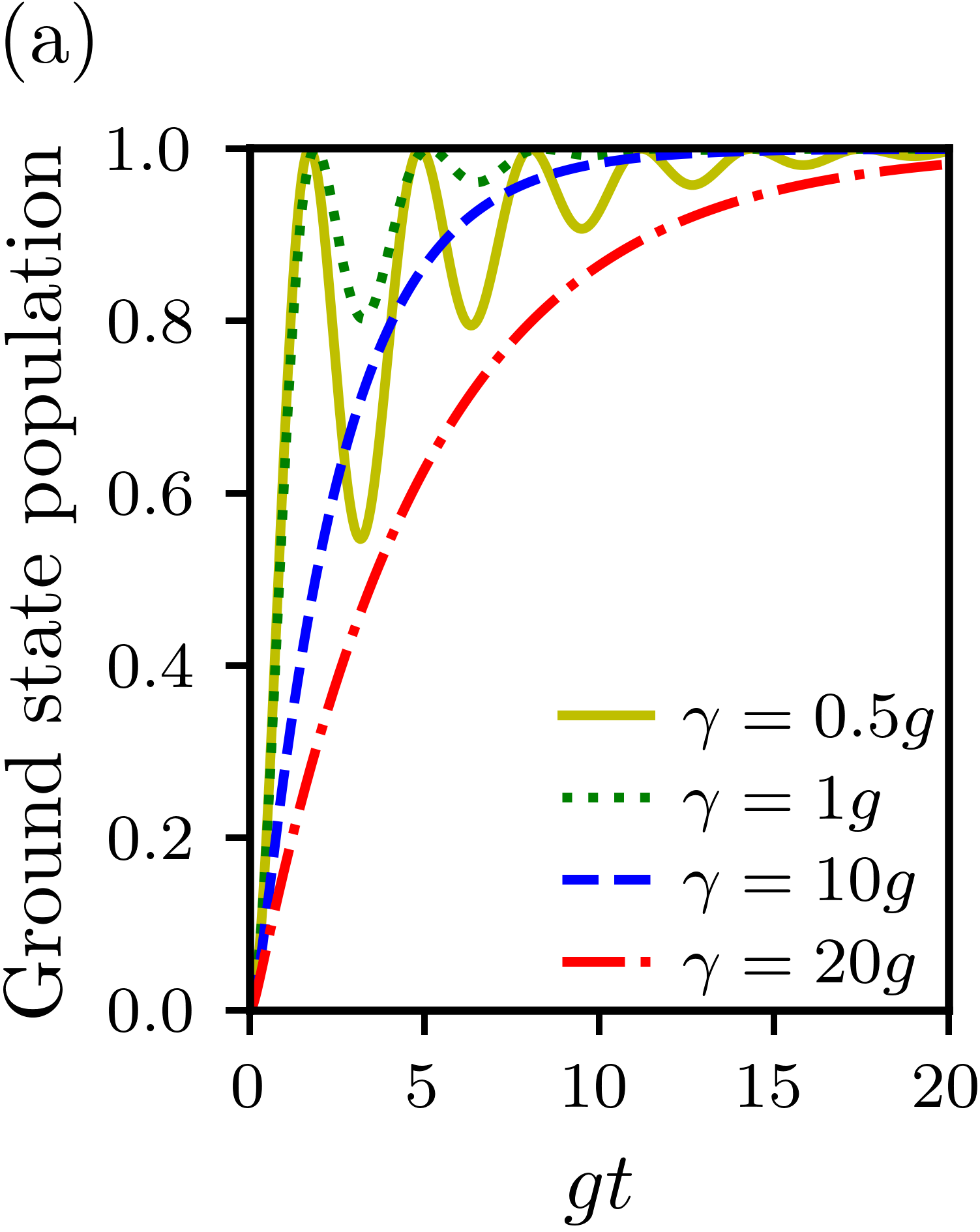}
\includegraphics{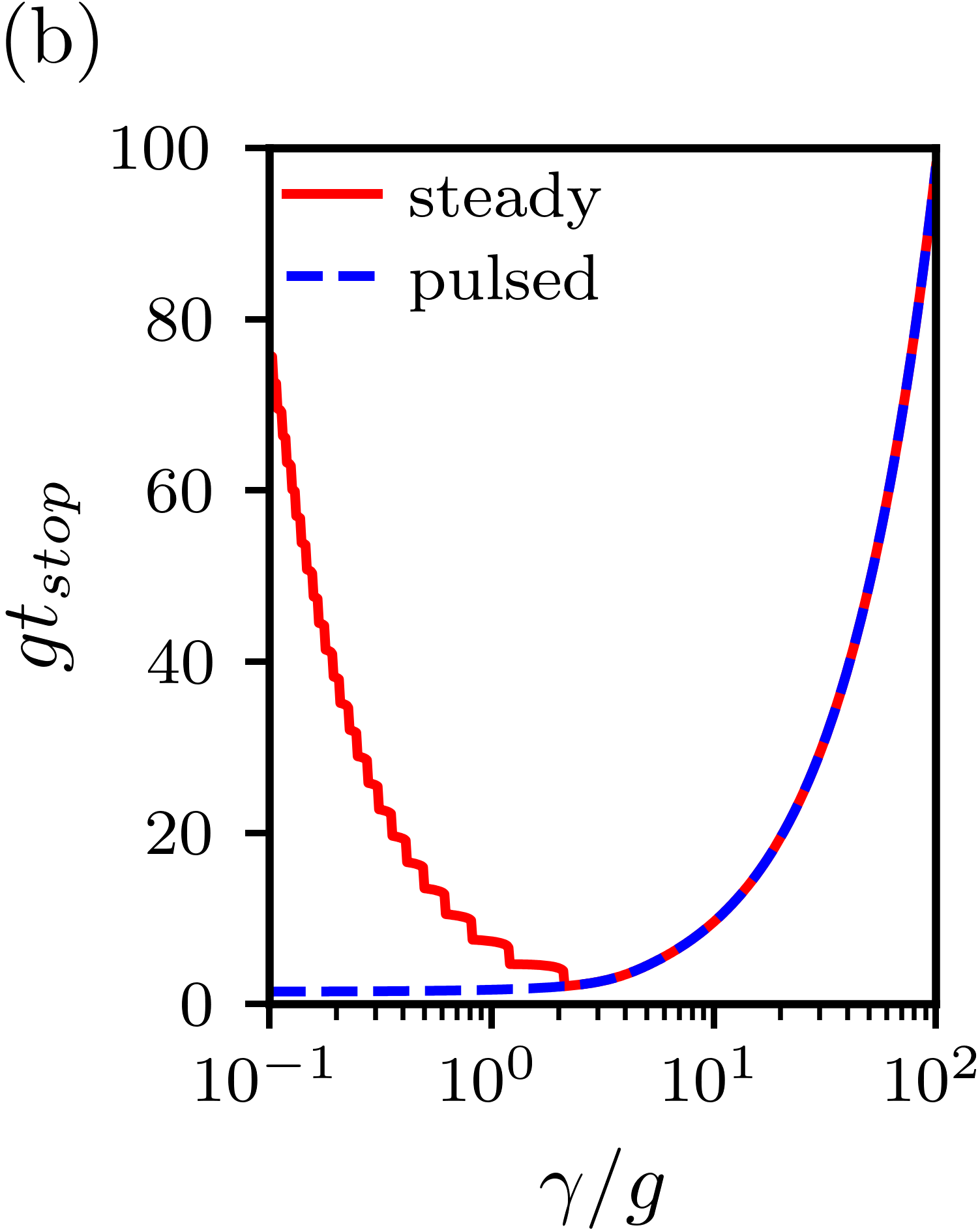}
\end{tabular}
\caption{Resetting process for the two-qubit model. The initial state of the main qubit is $\rho_M=\ketbra{e}{e}$ while the initial state of the auxiliary qubit is $\rho_A = \ketbra{g}{g}$. Panel (a) shows the ground state population of the main qubit as a function of the dimensionless time $gt$, for different values of the dissipation rate $\gamma$ - see legend. Panel (b) shows the dimensionless time $gt_{stop}$ needed to complete the reset as a function of the dissipation rate $\gamma/g$ for two reset approaches: the pulsed approach (dashed blue line) and the steady-state approach (solid red line).}
\label{DQ_geral}
\end{figure}

Now, to study the reset protocol for the models presented in Fig.\ref{New Schemes}(c)-(e), let us start considering the two-qubit model, Fig. \ref{New Schemes}(c), where we analyzed the dynamics of the ground state population of the main qubit for different values of the dissipation rate of the auxiliary qubit, given by $\gamma$. For this configuration, Fig. \ref{DQ_geral}(a) shows the ground state population versus $gt$. Looking at this panel, when considering the steady-state approach, we can notice that the best value is $\gamma=10g$. Unlike what would be expected, as the dissipation rate is increased regarding to the coupling strength between the two qubits, the system reset becomes ineffective, \textit{i.e.}, the resetting will take longer to occur, e.g., see the curve for $\gamma=20g$. Conversely, still in the same approach, if the dissipation rate is too small, it will take too long for the reset process to stabilize and finish, although the first peak of the dynamics occurs earlier, what is a good fact for the pulsed approach, as it happens for either $\gamma=0.5g$ or $\gamma=1g$ for instance.

In Fig. \ref{DQ_geral}(b) it is shown the time $g t_{stop}$ as a function of $\gamma/g$ for the two reset approaches. Starting with the pulsed approach, one can note that for values of $\gamma/g$ smaller than $2$, approximately, the $gt_{stop}$ does not depend on the decay rate of the auxiliary qubit, being the best times close to $g t_{stop} = 1.45$. On the other hand, when considering the steady-state approach, from Fig. \ref{DQ_geral}(a) we see that there is an optimum value for the reset time. From Fig. \ref{DQ_geral}(b), we numerically find that this optimum value ($gt_{stop} \sim 2.08$) occurs for $\gamma/g \sim 2.13$. For values greater than this ratio, the reset process is ineffective since the time $gt_{stop}$ increases as $\gamma/g$ increases. Still, for $\gamma/g <2$, the time also increases as the ratio $\gamma/g$ decreases. For instance, if we now consider the minimum ground state population of $0.995$, as required to perform realistic quantum computing, the best reset times achieved are $gt \sim 1.53 $ and $gt \sim 2.60$, for the pulsed and steady-state approaches respectively, where the parameter that optimizes the reset time for the latter approach is $\gamma/g \approx 2.59$.

Still regarding the two-qubit model, one can notice in Fig. \ref{DQ_geral}(a) that the dynamics for greater values of $\gamma$ do not oscillate, while the dynamics for smaller values of $\gamma$ do oscillate, as one can see from comparing the curves for $\gamma=20g$ and $\gamma=0.5g$. These oscillations for the smaller values of $\gamma$, which arise from the energy flux between the qubits, are the origin of the discontinuities existing in Fig. \ref{DQ_geral}(b) in the curve for the steady-state approach. In fact, in the beginning of the dynamics, the energy stored in the main qubit flows to the auxiliary qubit, which has a weak decay, \textit{i.e.}, $\gamma$ has the same magnitude or it is smaller than the coupling
strength $g$. Then, due to its small dissipation
rate, the energy that is transferred to the auxiliary qubit is able to flow back to the main one before it is completely dissipated, hence, the ground state population of the main qubit keeps recurring. This energy exchange between the qubits continues to occur and several oscillations in the dynamics of the ground state population take place until the reset process is over. It is worth noting that the number of oscillations depends on the value of $\gamma$, \textit{i.e.}, the more oscillations occurs, the longer the reset time is, and vice-versa.

\begin{figure}
\begin{center}
\begin{tabular}{cc}
\includegraphics{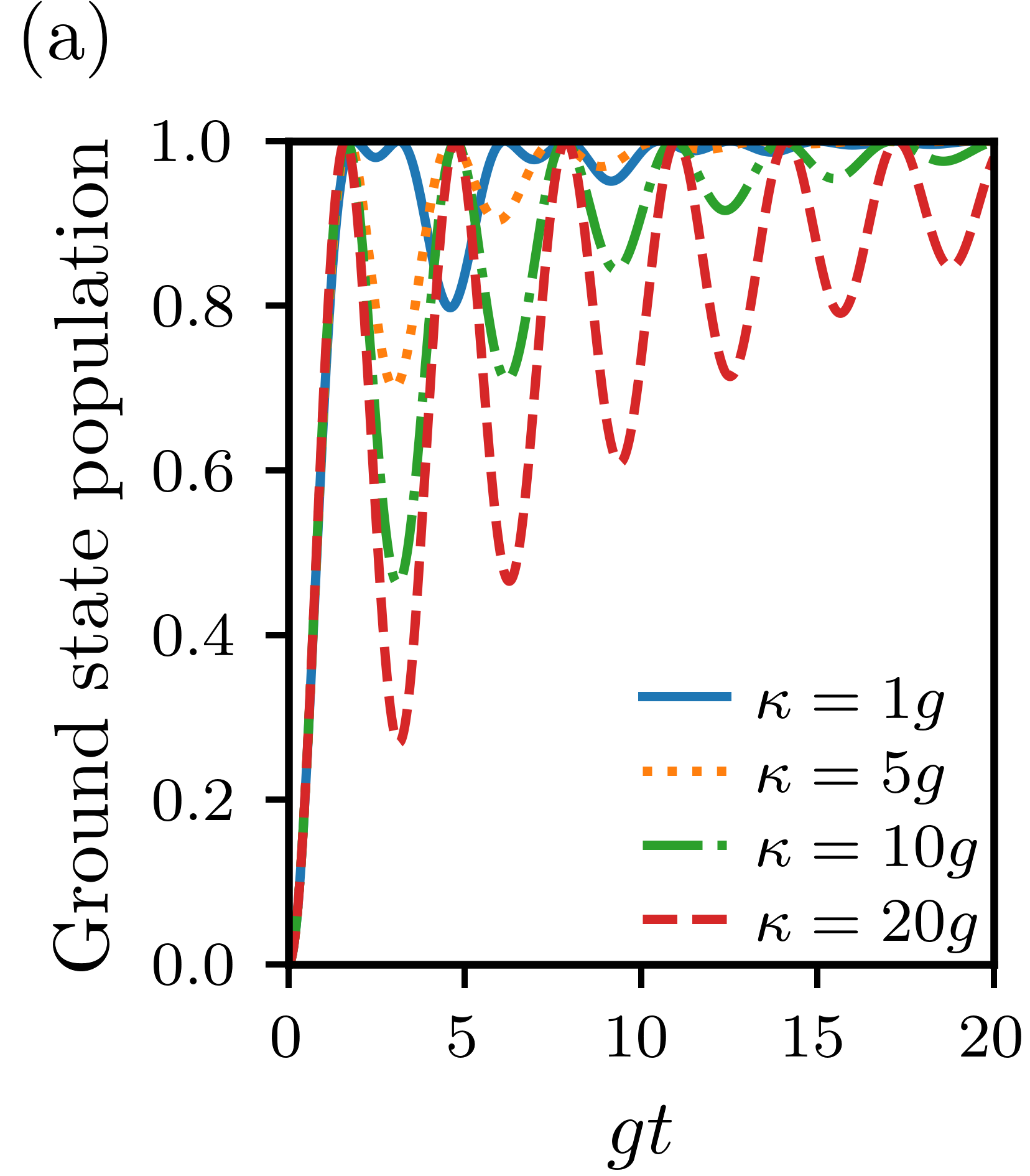}
\includegraphics{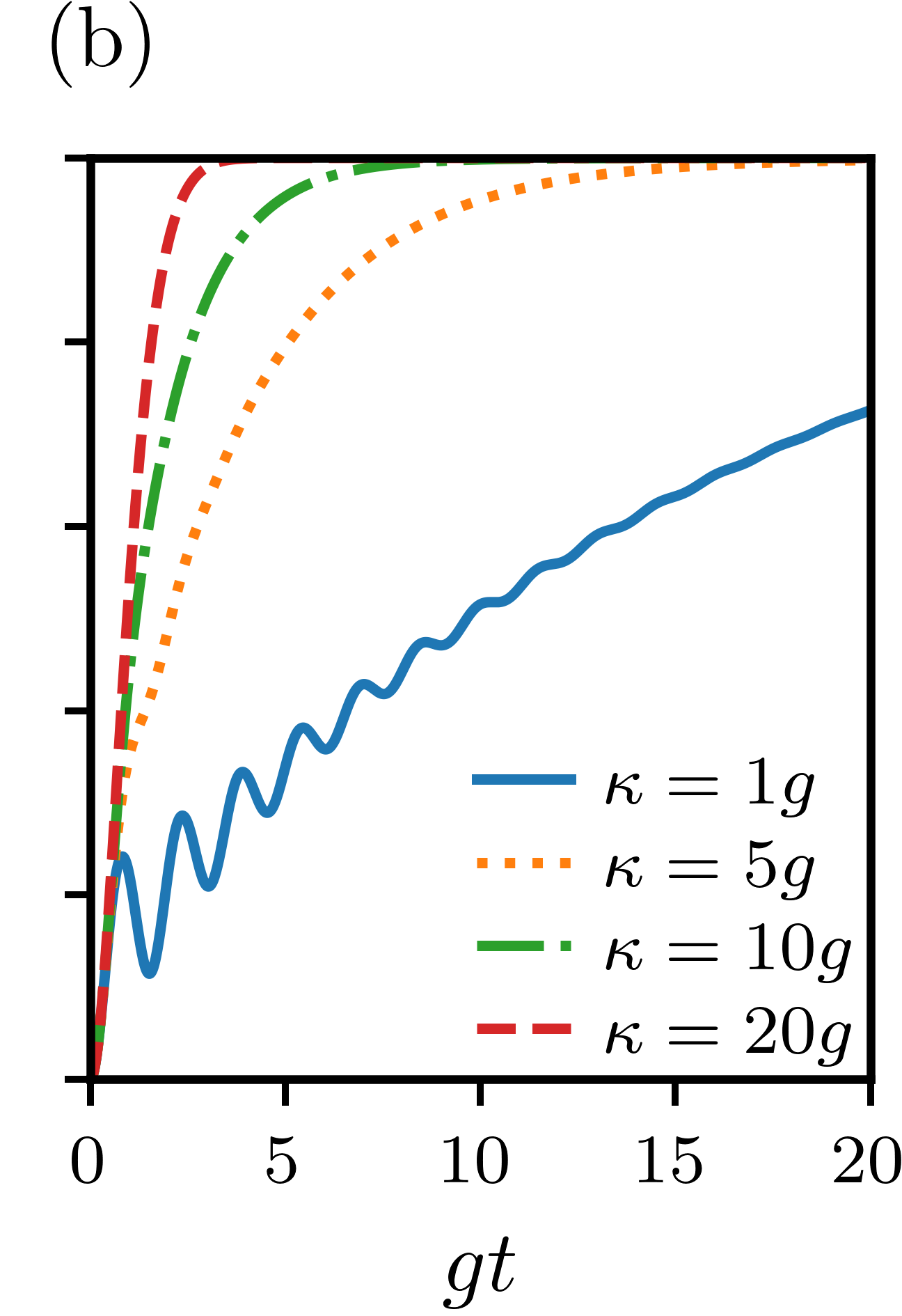}
\end{tabular}
\begin{tabular}{cc}
\end{tabular}
\begin{tabular}{cc}
\includegraphics{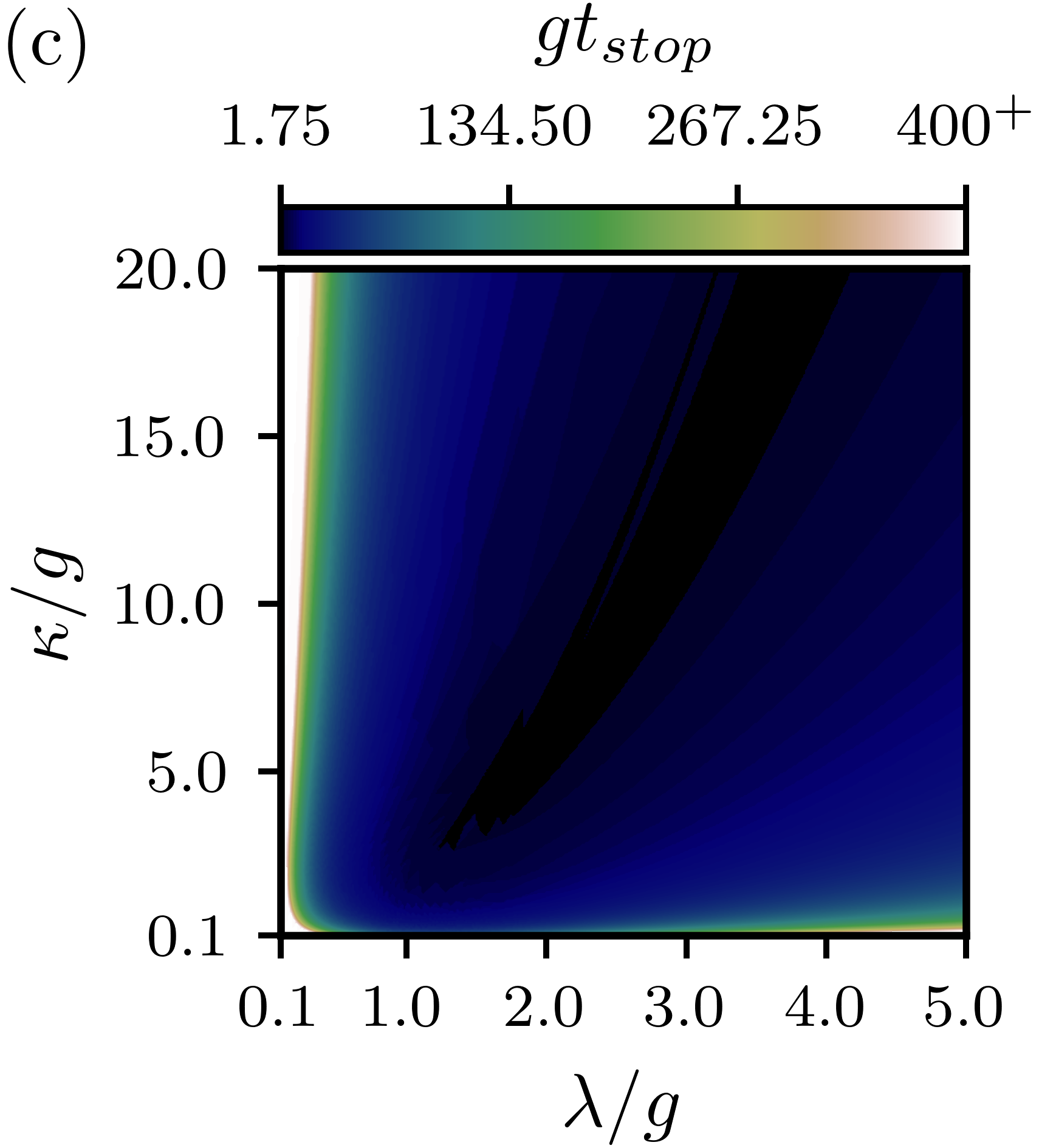}
\includegraphics{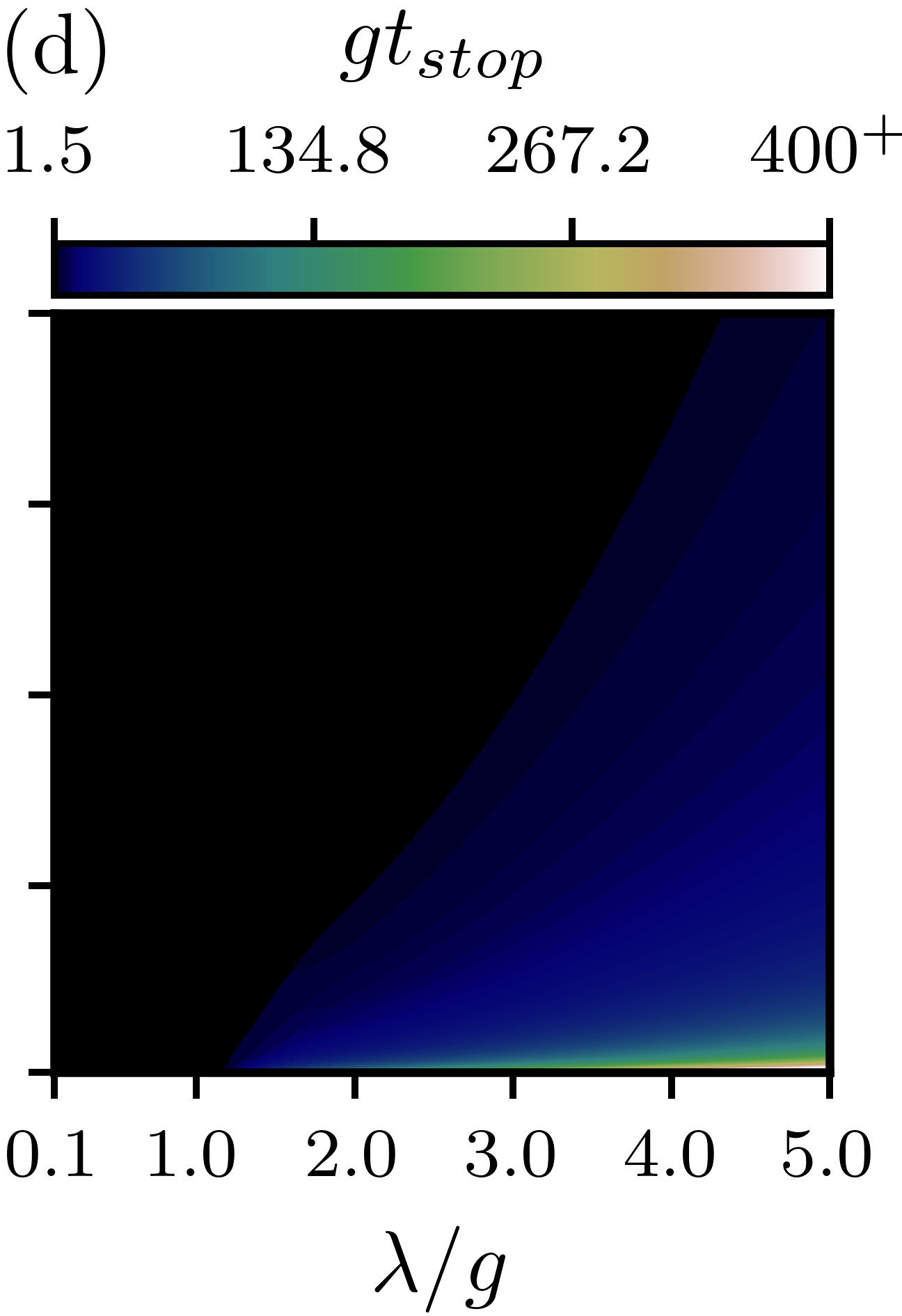}
\end{tabular} 
\caption{Resetting process for the two qubits-cavity model. In these simulations the initial state of the system is given by:  $\rho_M=\ketbra{e}{e}$, $\rho_A=\ketbra{g}{g}$, $\rho_c = \ketbra{0}{0}$, i.e., main qubit, auxiliary qubit, and cavity mode, respectively. The upper panels show the ground state population of the main qubit as a function of the dimensionless time $gt$ for two coupling strength values: (a) $\lambda=1g$ and (b) $\lambda=4g$, and different dissipation rates $\kappa$, as indicated in the legend. The lower panels show the $gt_{stop}$ optimized according to the parameters $\lambda/g$ and $\kappa/g$ for the two reset approaches: (c) steady state approach and (d) pulsed approach. In the blank areas of these last two panels, the time needed to reset the system is higher than the limit time $gt_{stop}=400$, although the reset still occur for longer times.}
\label{OUR_geral}
\end{center}
\end{figure}

For the two qubits-cavity model, represented in Fig. \ref{New Schemes}(d), we have three parameters, two from the Hamiltonian of Eq. (\ref{H our}) and one from the dissipation operator, which are respectively: the coupling strength $g$ between the two qubits, the coupling strength $\lambda$ between the auxiliary qubit and the cavity mode field; and the damping rate $\kappa$ of the electromagnetic (EM) field mode. Thus, by fixing the coupling $g$ as before, there will remain two parameters to be adjusted in order to optimize the resetting process. The results for this case are shown in Fig. \ref{OUR_geral}, where we plot in the upper panels the dynamics of the ground state population of the main qubit as a function of the dimensionless time $gt$ for $\lambda=1g$, Fig. \ref{OUR_geral}(a), and  $\lambda=4g$, Fig. \ref{OUR_geral}(b), in both cases considering different values for $\kappa$ -- see legend. Looking at these two panels, once again the previous discussion about steady-state population can be made, since there are values of $\kappa$ and $\lambda$ for which the dynamics either oscillates and takes a long time to stabilize with a ground state population above the predefined bound, or it does not fluctuate, but still takes time to reach the desired bound. Hence, for this approach, once more we can infer that there is a optimum reset time. On the other hand, for the pulsed approach, an initial analysis reveals that the ratio between $\lambda$ and $\kappa$ appears to determine when the reset occurs, since one can see from panel \ref{OUR_geral}(a) -- where we have the smaller value of $\lambda$ ($\lambda=1g$) -- that for all values of $\kappa$ the dynamics reach their first peaks fast whereas in panel \ref{OUR_geral}(b) -- where we have the bigger value of $\lambda$ ($\lambda=4g$) -- only the dynamics for the bigger values of $\kappa$ reach their peaks fast.

Turning now to the optimizations, the panel \ref{OUR_geral}(c) shows the $gt_{stop}$ as a function of $\kappa/g$ and $\lambda/g$ for the steady-state approach. From this panel we can see that the best reset times -- close to $gt_{stop}=1.75$ -- are found in a narrow range of the parameters $\lambda$ and $\kappa$ only. One can also note that there are blank areas in this panel, which represent the set of parameters for which it is not possible to reach the minimum ground state population of the main qubit ($p_{g}=0.98$) in a stable way and within the maximum stipulated time in our simulations ($gt_{max} =400$). For the pulsed approach, similar results can be seen in Fig. \ref{OUR_geral}(d), but now the set of parameters with the best resetting times is larger, being  the best times around $gt_{stop}\approx 1.5$. On balance, for the first two configurations (two-qubit model and two qubits-cavity model), we found that the best times for the pulsed approach are slightly shorter than the best times for the steady-state approach. In contrast, the disadvantage of former as compared to the latter approach is the need for a high level of precision in controlling the duration of the time pulse. In fact, since in the pulsed approach we have to stop the time evolution in a very precise time, we need an experimental precision of the order of $1/g$, meaning that even a small inaccuracy in the stopping time of the dynamics will cause the system state to be far from the ground state, thus introducing errors and making the resetting process ineffective. If we consider, as before, the minimum ground state population of $0.995$, the best reset times do not change in both approaches. The changes in the results happen just in the set of parameters that reaches the required population, \textit{i.e.}, the same best reset times are achieved, but a smaller set of parameters achieves the required population during these best times.

%\begin{figure}[tb]
    %\centering{}

    %\includegraphics[scale=0.5]{OUR_otimizado.png}
    %\begin{picture}(0,0)
    %\put(-65,45){\includegraphics[width=4cm,height=2.5cm]{OUR_otimizado_inset.png}}
   % \end{picture}
    
  %  \caption{OUR: Dynamic for the main qubit coupled to an auxiliary qubit  which, in turn, is coupled to a dissipative bosonic mode, as a function of the dimensionless time $gt$, with $g=1$, to study the resetting process. Ground state population  to the qutrit The system initial state is $\rho_M=\ketbra{e}{e}$, the reservoir initial state $\rho_R=\ketbra{g}{g}$ and the initial state of the cavity mode is $\rho_F = \ketbra{0}{0}$. Each curve shows the better optimization of the parameters, for $\lambda=1, 2, 3, 4$ and $15$ the optimum values for $\kappa$ are, respectively, $\kappa=3, 6, 14, 22$ and $260$ - see legend. }
 %   \label{OUR_otimizado}
%\end{figure}

%\subsection{Sistema da IBM}

\begin{figure*}
\begin{center}
\begin{tabular}{cc}
\includegraphics{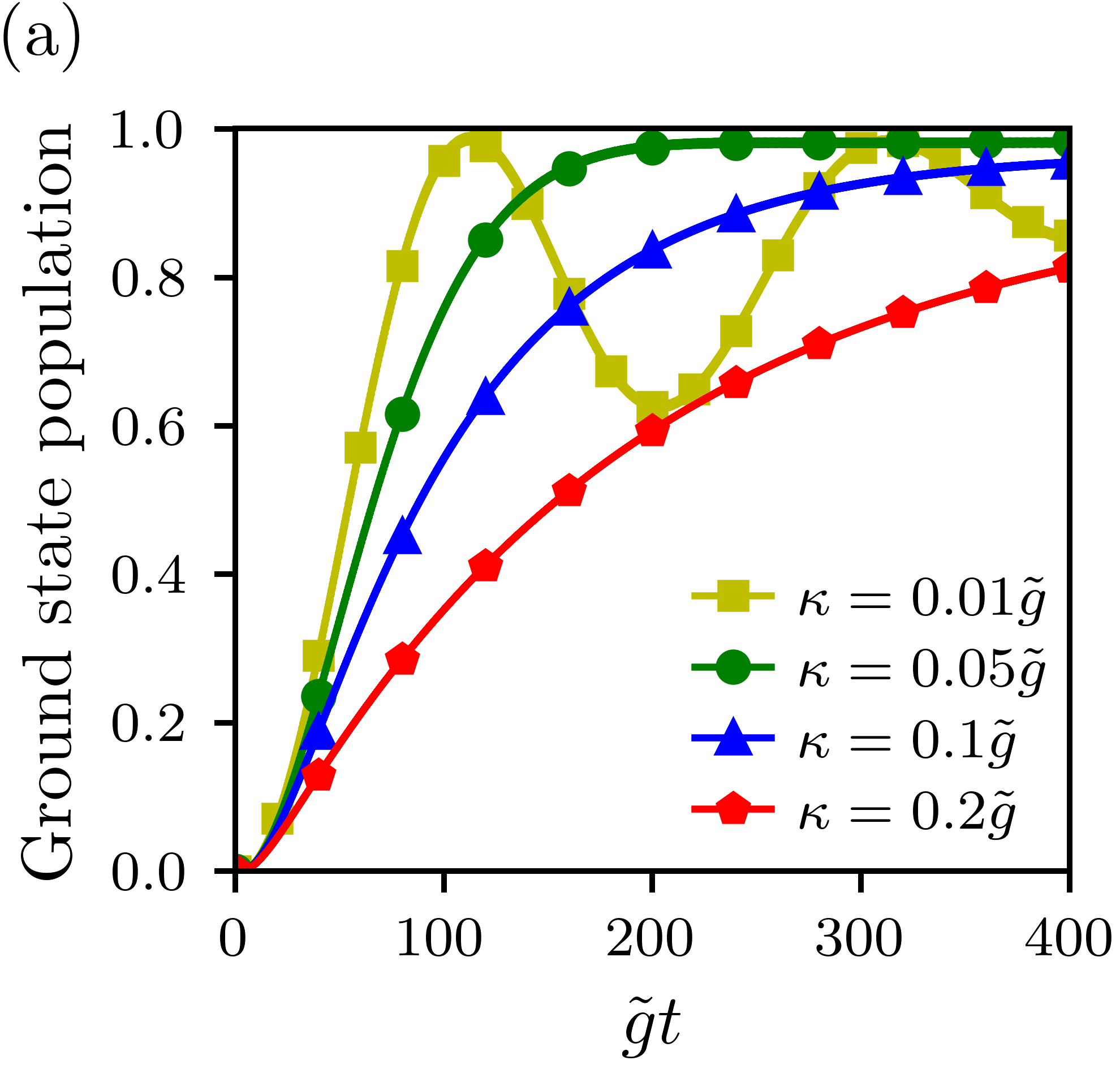}
\includegraphics{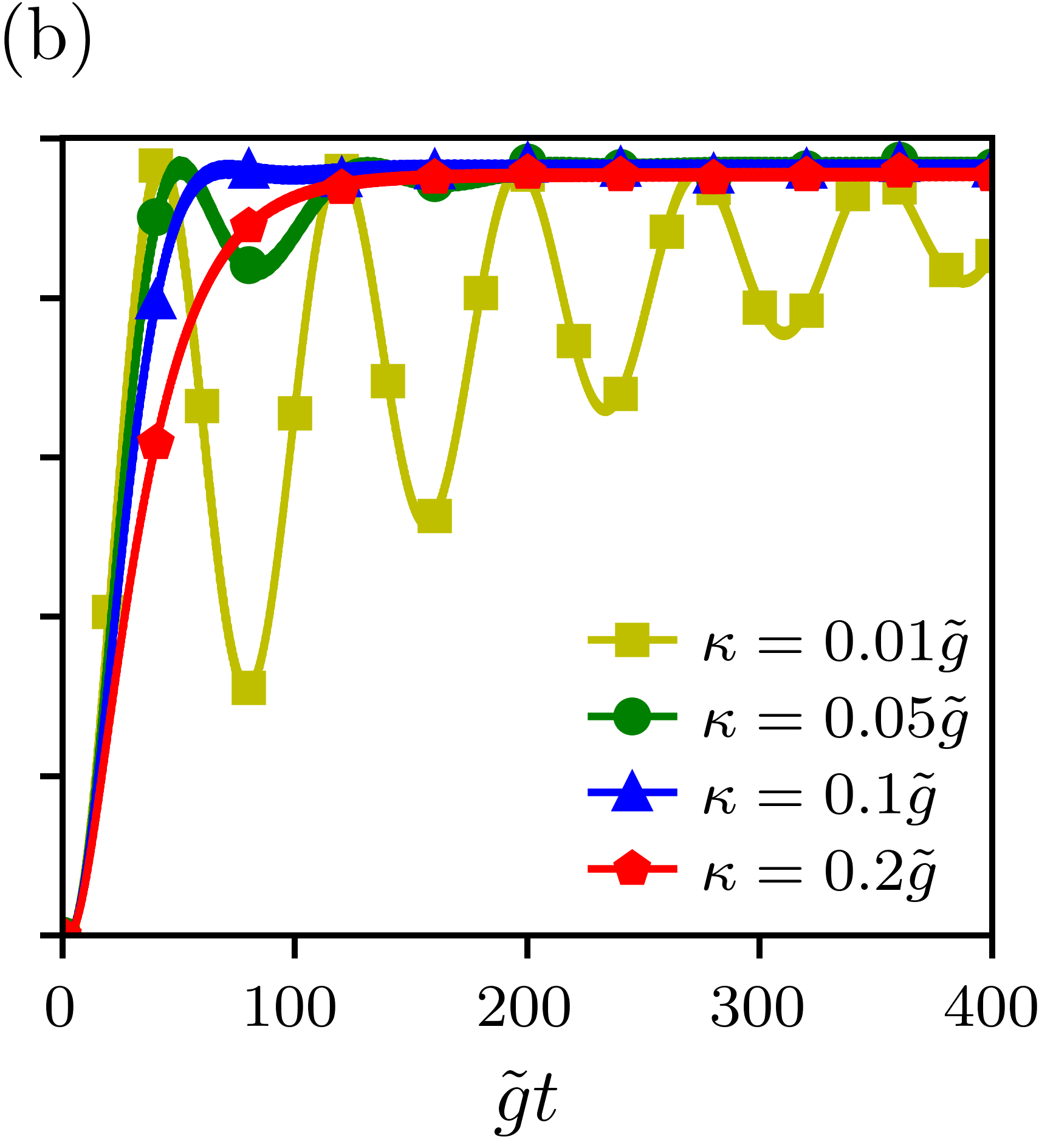}
\includegraphics{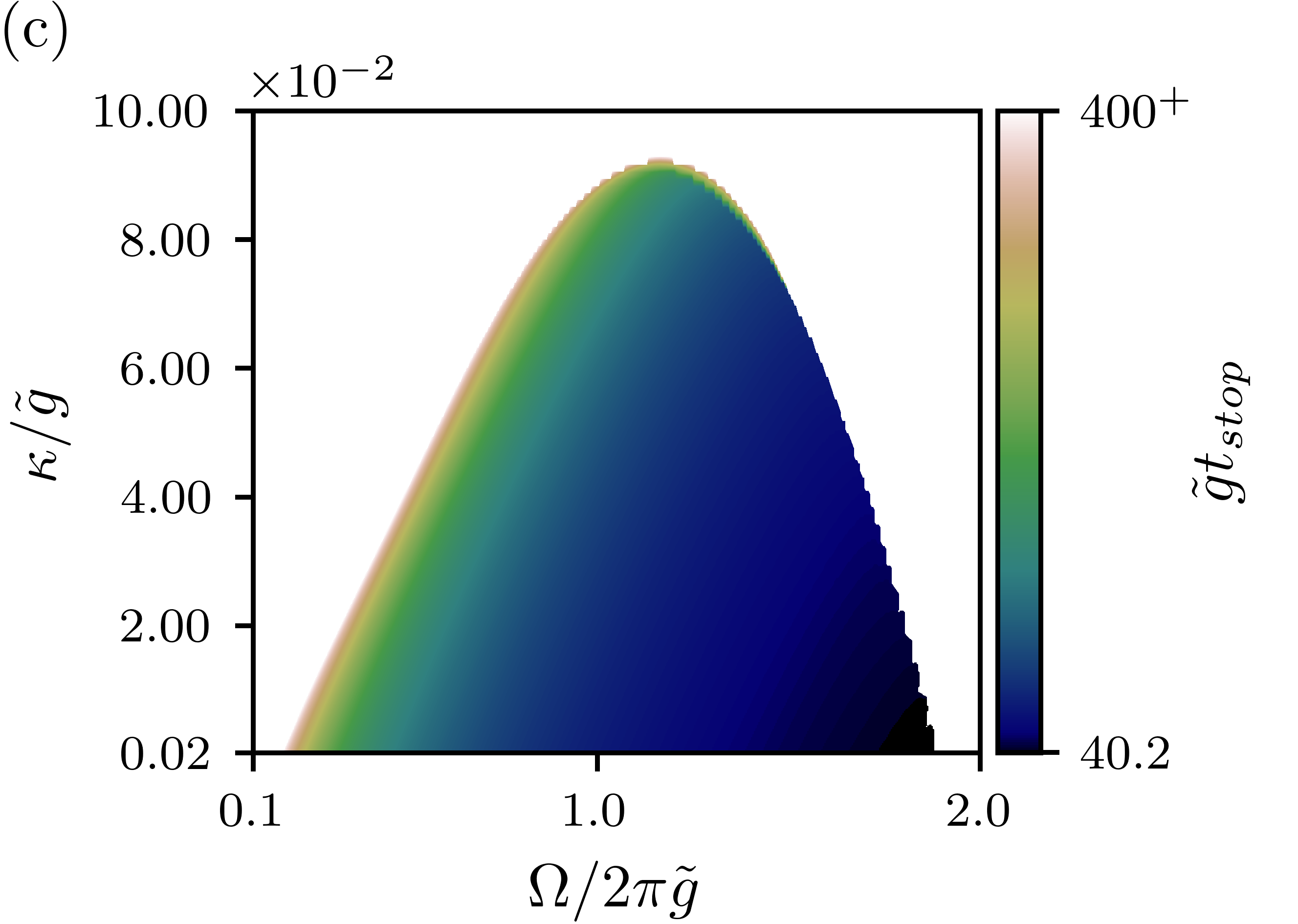}
%\begin{picture}(0,0)
    %\put(-89,30){\includegraphics[width=2.9cm,height=2.5cm]{IBM_otimizado_inset.png}}
%\end{picture}
\end{tabular} 
\caption{Resetting process in the IBM model studied in \cite{IBM}. In these simulations the initial state of the main system is $\rho_M=\ketbra{f}{f}$ while the initial state of the cavity mode is $\rho_c = \ketbra{0}{0}$. We considered different dissipation rates $\kappa/\Tilde{g}$, as indicated in the legend. Panels (a) and (b) show the ground state population of the main qubit as a function of the dimensionless time $\Tilde{g}t$. The driven strength in the plots are, respectively, $\Omega/\Tilde{g}=0.6\pi$ and $\Omega/\Tilde{g}=1.6\pi$. The panel (c) shows the $\Tilde{g}t_{stop}$ as a function of the parameters $\Omega/2\pi \Tilde{g}$ and $\kappa/\Tilde{g}$. In the white area the time needed to reset the system is higher than the maximum stipulated time $\Tilde{g}t_{stop}=400$.}
\label{IBM_geral}
\end{center}
\end{figure*}

The IBM model \cite{IBM} is the last configuration studied here. % and the Fig. \ref{IBM_geral} contains its results. 
Again, since this model was developed aiming the pulsed approach, we just searched for its optimization, \textit{i.e.}, we did not consider the steady-state approach for this model. As said before, this model occurs in two steps. Firstly the population of the excited state $\ket{e}$ is transferred to the auxiliary state $\ket{f}$ via auxiliary pulse. Since the intensity of the pulse modifies either the time spent in this step and the possible maximum transferred population, we consider here the same pulse that was used in \cite{IBM}, which takes $75$ $ns$ to transfer the population to the state $\ket{f}$, approximately. Once the main system is in the state $\ket{f}$, the second step starts coupling the system to the dissipative cavity mode in order to bring it to the desired ground state. Hence, to optimize the second step of the approach, and consequently the entire process, we searched for the parameters that make the dissipative dynamics of the transition $\ket{g} \leftrightarrow \ket{f}$ faster. 

Proceeding as in the previous cases, in Figs. \ref{IBM_geral}(a)-(b) we show the dynamics of the ground state population of the main qubit as a function of the dimensionless time $\Tilde{g}t$. Remembering we are considering the pulsed approach only, from Fig. \ref{IBM_geral}(a) we can see that, for a given value of $\Omega/\Tilde{g}$, very large values of $\kappa/\Tilde{g}$ do not improve the resetting process, \textit{i.e.}, the higher $\kappa/\Tilde{g}$ the longer the reset will take. However, by increasing  $\Omega/\Tilde{g}$, as shown Fig. \ref{IBM_geral}(b), we can notice that the reset becomes faster, once, as we can see from the comparison of the two panels, for a fixed $\kappa/\Tilde{g}$, the first peak of the ground state population happens earlier for $\Omega/\Tilde{g}= 1.6\pi$.

Looking for the optimization process, panel \ref{IBM_geral}(c) shows two regions. The colored region shows the $\Tilde{g}t_{stop}$ as a function of $\kappa/\Tilde{g}$ and $\Omega/2\pi \Tilde{g}$. Note from this figure that the shortest time spent in the second step of the reset is $\Tilde{g}t_{stop}\approx 40.2$ ($\sim 95.5$ $ns$, when considering $\Tilde{g}/2\pi=67$ MHz as in \cite{IBM}), which results in a total resetting time, \textit{i.e.}, considering the two steps of the approach, of $170.5$ $ns$, thus decreasing almost $20\%$ the resetting time, as compared to the result achieved in Ref. \cite{IBM}, that is $210$ $ns$. The other region is the blank area, which represents again the set of parameters whose resetting times are longer than the limit of $\Tilde{g}t=400$. For these parameters the resetting delays due to one of following reasons: (i) the dynamics for a given $\Omega/2\pi \Tilde{g}$ and $\kappa/\Tilde{g}$ naturally take to long -- this behavior happens for small values of $\Omega$ mostly, \textit{e.g.}, see the curve for $\kappa=0.2\Tilde{g}$ at Fig. \ref{IBM_geral}(a); or (ii) as said before, the efficiency of the transition $\ket{g} \leftrightarrow \ket{f}$ depends on the intensity of the pulse, therefore as the value of $\Omega/2\pi \Tilde{g}$ increases, the smaller the efficiency of this transition is, so that the interaction between the qubit and the cavity has to be kept for longer until the bound population of $p_g=0.98$ is reached, and occasionally this time surpasses the predefined limit of $\Tilde{g}t=400$.

An important question that could rise is the impact for quantum computing of the lossy effects induced by the auxiliary systems in the dynamics, because they can harm the performance of the device. When the main qubit is not working, some methods can be applied to reduce the crosstalk due to the interaction of the main qubit, and consequently the reset components, with the other working qubits.  For example, following \cite{Xu2020} and \cite{Hu2023}, the couplers present in the architecture of the computer can be used to reduce such crosstalk and even effectively cancel the interactions between two specific components, which are the main qubit and other working qubits present in the architecture of the quantum device. On the other hand, when the main qubit is actively involved in calculation tasks, it is possible to set the frequency of the auxiliary qubit far off the resonance, which implies in a large detuning between the main and auxiliary qubits that cancels the interaction between them. Therefore, either during the resetting process and the computing step, it is possible to avoid the undesired crosstalk and strong induced decays of the qubits.

For instance, consider the coupling strengths close to the ones in Ref. \cite{Hu2023}, which are around $g/2\pi = 10$ MHz. The Purcell decay time for the two-qubit model, Fig. \ref{New Schemes}(c), is expressed as $T_{\text{Purcell}} = (\Delta / g)^2\gamma^{-1}$, with $\Delta$ being the detuning between the main and the auxiliary qubits during the computing step \cite{Houck2008}. Thus, assuming that the effective decay induced by the auxiliary qubit is of the order of the natural decay of the main qubit, that is of the order of $\ 50 $ $\mu s$ \cite{IBM, Burnett2019},  we note that detunings of the order of $1.5$ GHz up to $2.5$ GHz will be required. For instance, for the steady-state approach, it is possible to achieve $T_{\text{Purcell}} = 50$ $\mu s$ using detunings close to $2.5$ GHz and sets of parameters ($\gamma/g \sim 0.5$) that result in reset times of $20$ $ns$ approximately. On the other hand, for the pulsed approach, it is possible to achieve $T_{\text{Purcell}} = 50$ $\mu s$ considering detunings smaller than $1.5$ GHz, even if the optimum sets of parameters is used ($\gamma/g \sim 0.1$). Noteworthy, the required detunings are feasible to guarantee these reasonable decay times \cite{Hutchings2017, Houck2008, Bronn2015}. Now, for the two qubits-cavity model, Fig. \ref{New Schemes}(d), it is necessary taking into account the effective decay rate of the component coupled to the main qubit. In the second model studied here, the auxiliary qubit has an effective decay rate, due to its resonant coupling to the dissipative cavity mode, which is given by $\Gamma_{\text{eff}} = \lambda^2/\kappa$ \cite{Prado2009, TRZ}, such that the Purcell decay time of the main qubit has to be computed as $T_{\text{Purcell}} = (\Delta / g)^2 \left(\lambda^2/\kappa\right)^{-1}$. In this context, it is possible to achieve the same Purcell decay times from before, $T_{\text{Purcell}} = 50$ $\mu s$, considering for the pulsed approach detunings of $1.5$ GHz approximately and the optimum sets of parameters, while the steady-state approach requires detunings of $2$ GHz approximately and sets of parameters that result in reset times smaller than $100$ $ns$. Additionally, if one considers a coupling strength similar to the one used in the IBM paper \cite{IBM}, which is $\Tilde{g}/2\pi = 67$ MHz, it is possible to achieve Purcell decay times around $1$ $\mu s$ considering detunings smaller than $3$ GHz, which are still feasible, and sets of parameters that result in reset times of $100$ $ns$ approximately.

\section{Conclusion}

%\textcolor{magenta}{In this work we provided a comprehensive study for the time reset process in quantum computers. We demonstrated that it is not enough to achieve a faster reset time by just increasing the coupling strength between the reset components of a quantum computer, and thus, a deeper analysis is needed to consider all the relevant parameters involved.} 
In this work we have analyzed the optimization of the resetting time of superconducting qubits in three models: (i) two-qubit model, where the main qubit is coupled to a second dissipative one, (ii) two qubits-cavity model, composed by two interacting qubits with coupling strength $g$, one being the main qubit and the other being the auxiliary qubit, which is coupled to a dissipative cavity mode, and (iii) IBM model, where a qubit and an auxiliary level are coupled  to a dissipative bosonic mode. To study the resetting times in the first two models, which are analyzed here for the first time, we consider two different situations: the steady-state and the pulsed approaches. The model (iii), which was designed to operate with the pulsed approach only, was introduced first in Ref. \cite{IBM}. After the optimizations we were able to reduce the resetting time achieved in \cite{IBM} at about $20\%$ for their model, as they achieved  $210$ $ns$ for the resetting time and here, using the same coupling strength $\Tilde{g}/2\pi=67$ MHz, we could achieve the resetting time of $170.5$ $ns$. Furthermore, we note that this best resetting time of the IBM model is two orders of magnitude larger than the best resetting times of the two other models analyzed here, which are lesser than $5$ $ns$, where we are considering again the same coupling strength as before. Concerning the models (i) and (ii), in both of them the pulsed approach is slightly faster than the steady-state approach. However, it is necessary to keep in mind a possible disadvantage of the first approach compared to the latter, as pulsed approach requires high experimental accuracy in its execution to control the system dynamics. Finally, it is important to notice that one could argue that it could be possible to achieve shorter reset times in practice just employing even stronger couplings. However, the couplings between the subsystems can not be increased arbitrarily, once we could end up in the ultra-strong or deep-strong coupling regimes \cite{Yoshihara2017,Mercurio2022}, which are the regimes in which the interaction energies between the subsystems are of the same order or even higher than the free energies of the subsystems. In these regimes the ground state of the system is an entangled state between the subsystems \cite{Shitara2021,Beaudoin2011}, with virtual photons that make it difficult to prepare the main qubit in its ground state alone.

%Finally, in this work, during the optimization of the reset protocol for the IBM model, we showed its inherent limitations with regard to the possible values of $\Omega$ necessary for the protocol to work.

%%%%In this work we analyze three protocols for resetting the information, namely (i) two qubits, where the main qubit is coupled to a second dissipative qubit with a dissipation rate $\gamma$, (ii) a qubit coupled to a second one with interaction strength $g$ and this second qubit coupled to a dissipative cavity mode of damping rate $\kappa$, and (iii) a qubit and an auxiliary level that is coupled  to a dissipative bosonic mode of damping rate  $\kappa$. One might think that to decrease the time to reset the information it would be enough to increase the dissipation rate of the reservoir that absorbs the qubit excitation encoding the information. Our results, however, show that this is not the case: in all three setup analyzed here, the time for the effective reset of the information will depend on the relationship between the parameters $g$, $\gamma$, $\lambda$ and $\kappa$. Through a detailed analysis we were able to show for which set of parameters the reset of information occurs in the shortest time possible, and, in addition, makes the dynamic of the qubit to stabilize in an asymptotic state with no recurrence of information to the system, thus ensuring the effective reset of the information.

\section*{Acknowledgments }

We acknowledge financial support from the Brazilian agencies: Coordenação de Aperfeiçoamento de Pessoal de Nível Superior (CAPES), Financial code 001, National Council for Scientific and Technological Development (CNPq), Grants No. 311612/2021-0 and 301500/2018-5, São
Paulo Research Foundation (FAPESP) Grants No. 2019/11999-5, 2019/13143-0, 2022/10218-2, and No. 2021/04672-0, and Goiás State Research Support Foundation (FAPEG).
This work was performed as part of the Brazilian National Institute
of Science and Technology for Quantum Information (INCT-IQ/CNPq) Grant
No. 465469/2014-0.

\bibliography{bib1.bib}

\end{document}